\documentclass[sigconf]{acmart}
\usepackage{diagbox}
\usepackage{booktabs}
\usepackage{makecell}
\usepackage{subcaption}
\AtBeginDocument{%
  }

\setcopyright{acmlicensed}
\copyrightyear{2025}
\acmYear{2026}
\acmDOI{XXXXXXX.XXXXXXX}
\acmConference[ACM Conference on Intelligent User Interfaces '2026]
  {March 23--26, 2026}
  {Paphos, Cyprus}

\begin{document}

\title[Unmasking Hiring Bias]{Unmasking Hiring Bias: Platform Data Analysis and Controlled Experiments on Bias in Online Freelance Marketplaces via RAG-LLM Generated Contents}


\author{Wugeng Zheng}
\email{zheng.wug@northeastern.edu}
\orcid{0009-0000-1282-2069}
\affiliation{%
  \institution{Northeastern University}
  \city{Boston}
  \state{Massachusetts}
  \country{USA}
  \postcode{02115}
}

\author{Guohou Shan}
\email{g.shan@northeastern.edu}
\orcid{0000-0002-3268-6348}
\affiliation{%
  \institution{Northeastern University}
  \city{Boston}
  \state{Massachusetts}
  \country{USA}
  \postcode{02115}
}



\renewcommand{\shortauthors}{Zheng et al.}

\begin{abstract}
Online freelance marketplaces, a rapidly growing part of the global labor market, are creating a fair environment where professional skills are the main factor for hiring. While these platforms can reduce bias from traditional hiring, the personal information in user profiles raises concerns about ongoing discrimination. Past studies on this topic have mostly used existing data, which makes it hard to control for other factors and clearly see the effect of things like gender or race. To solve these problems, this paper presents a new method that uses Retrieval-Augmented Generation (RAG) with a Large Language Model (LLM) to create realistic, artificial freelancer profiles for controlled experiments. This approach effectively separates individual factors, enabling a clearer statistical analysis of how different variables influence the freelancer project process. In addition to analyzing extracted data with traditional statistical methods for post-project stage analysis, our research utilizes a dataset with highly controlled variables, generated by an RAG-LLM, to conduct a simulated hiring experiment for pre-project stage analysis. The results of our experiments show that, regarding gender, while no significant preference emerged in initial hiring decisions, female freelancers are substantially more likely to receive imperfect ratings post-project stage. Regarding regional bias, a strong and consistent preference favoring US-based freelancers shows that people are more likely to be selected in the simulated experiments, perceived as more leader-like, and receive higher ratings on the live platform.
\end{abstract}



\maketitle

\section{Introduction}

Online freelance marketplaces are increasingly becoming a vital component of the global labor market after the COVID-19 crisis \cite{international_labour_office_world_2021}. In Europe and the United States, 20 to 30 percent of the working-age population engage in some form of independent work \cite{noauthor_independent_nodate}. These platforms provide workers with flexibility to choose projects that fit their skills while giving employers access to a diverse talent \cite{kittur_future_2013}\cite{kittur_crowdforge_2011}. In theory, this task-oriented model should boost a more equally competitive environment, where professional capabilities, rather than personal background, serve as the core criterion for evaluation \cite{wu_future_2023}.

However, while online platforms can mitigate the direct discrimination present in traditional face-to-face recruitment \cite{lippens_state_2023}\cite{kroll_discriminatory_2021}, it remains unclear whether they have truly achieved the goal of eliminating bias. User profiles on these platforms contain personally identifiable information, leading to both unconscious and conscious biases during candidate evaluations \cite{hangartner_monitoring_2021}. While prior studies have confirmed the presence of gender and racial bias on online freelance platforms using statistical regression \cite{hannak_bias_2017}\cite{foongeureka_women_2018}, their methodologies face critical limitations: Analyzing observational data extracted from sources like freelancer profiles and ratings makes it difficult to control for confounding factors, thus hindering the precision of the findings. The conventional solution, collecting massive datasets to statistically account for these variables poses significant data acquisition challenges.

To more rigorously examine the bias on freelancer platforms, a new experimental methodology designed for strict variable control is proposed. A new method that employing Retrieval-Augmented Generation (RAG) to prompt a Large Language Model (LLM) to generate highly relevant experimental data, including freelancer profiles. By inviting participants to participate in the controlled experiments, we can thoroughly examine the biases on the freelancer platforms. Figure \ref{fig:Workflow} shows the workflow of our method. The key methodological contributions of this research are: 

\begin{figure}
    \centering
    \includegraphics[width=0.99\linewidth]{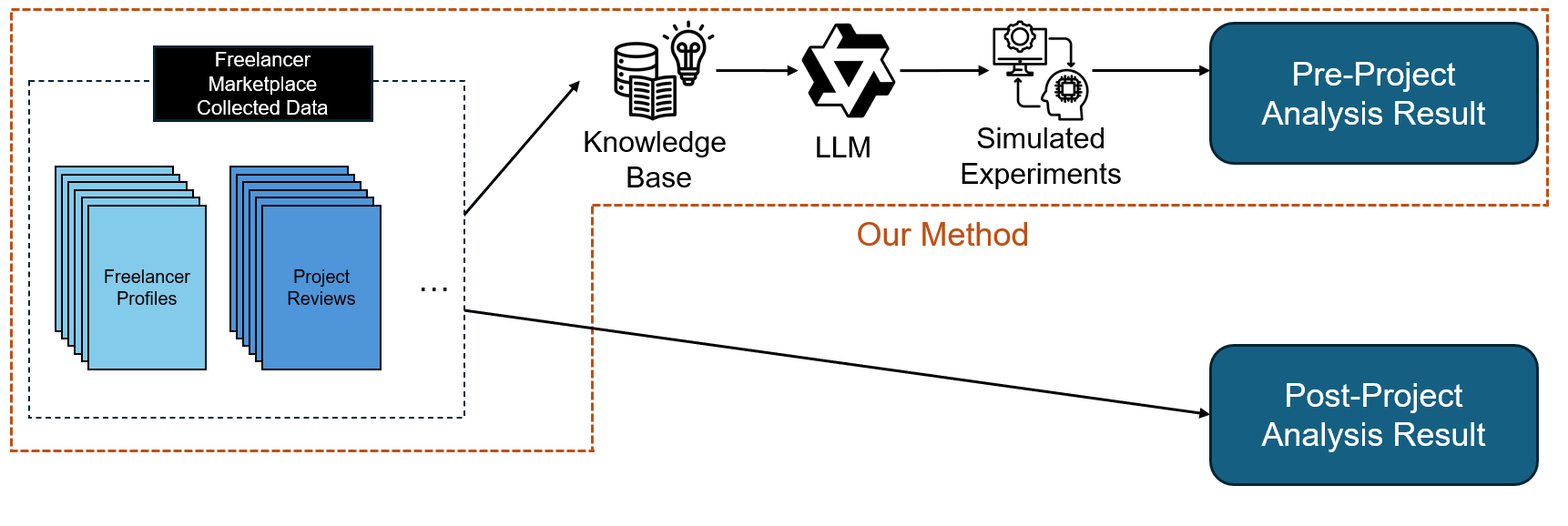}
    \caption{Workflow}
    \label{fig:Workflow}
\end{figure}


\begin{enumerate}
    \item \textbf{Controlled Experiments via Synthetic Data:} The usage of RAG-LLM generating synthetic data for controlled experiments allows the precise isolation of variables, overcoming the challenge of confounding factors present in traditional observational data.
    \item \textbf{Multi-Stage Bias Analysis:} The multi-stage analysis of bias by examining both the initial hiring phase through a user study and the post-project evaluation phase using real-world data provides a more complete view than studies limited to post-project data, revealing how different biases manifest at different stages of the engagement process.
\end{enumerate}

\section{Related Work}
\subsection{Discrimination in Online Marketplaces}
With the development of internet technology and the evolution of work models, the global economy has exhibited the rapid rise of the "gig economy"\citep{hoque_how_2015}. This new economic model, characterized by its flexibility and decentralization, has reshaped the organizational forms of the traditional workforce. Online freelance marketplaces, such as Upwork, Fiverr, and Freelancer.com, are central platforms of the gig economy. By connecting clients and freelancers on a global scale, these platforms break down geographical barriers and provide unprecedented opportunities for both parties \cite{graham_digital_2017}\cite{mierina_rise_2024}\cite{burns_relational_2024}. This mechanism enhances matching efficiency through information transparency and market competition \cite{guo_matching_2025}. In theory, it should allow skills to be the decisive factor, thereby reducing the potential for biases that may exist in traditional hiring processes \cite{veltri_impact_2023}.

However, the existing studies have found that discrimination continues to be a significant challenge. Profiles on these platforms often contain personally identifiable information, such as names and photographs, which creates opportunities for both explicit and implicit bias. Hannak \cite{hannak_bias_2017} found that on platforms like TaskRabbit and Fiverr, workers perceived as Black received significantly lower ratings than their peers from other ethnic groups, while female workers received a comparatively lower number of reviews. Edelman \cite{edelman_racial_2017} argued that even on sharing economy platforms with relative information transparency, such as Airbnb, consumer discrimination persists. In the absence of credible, low-cost information on individual quality, consumers may tend to group characteristics like race as a proxy, resulting in statistical discrimination against minorities. A field experiment conducted by Chan \cite{chan_hiring_2018} in an online labor market found that hiring preferences can sometimes favor female applicants, with employers in certain contexts showing a greater inclination to hire women.

\subsection{Machine Learning and Large Language Models}
Traditionally, methods for studying online discrimination have primarily relied on data scraping and econometric analysis, like works of Hannak \cite{hannak_bias_2017} and Chan \cite{chan_hiring_2018}. However, such approaches have an inherent limitation: The complexity of real-world data makes it challenging for researchers to control for all potential confounding variables. For instance, a candidate's individual qualifications, such as skills and experience, are often intertwined with their demographic identity (e.g., race or gender), complicating efforts to precisely identify the true source of bias.

The complexity of real-world data makes it challenging for researchers to control for all potential correlation variables. For instance, a candidate's individual qualifications, such as skills and experience, are often intertwined with their demographic identity (e.g., race or gender), complicating efforts to precisely identify the true source of bias. Some studies, like Liu et al.  \cite{liu_biaseye_2024} used individual screening preferences modeling by providing a synthetic profile methodology to process controlled experiments, to identify and mitigate biases

Recently, generative artificial intelligence, particularly Large Language Models (LLMs) has been used to understand the user activity on digital platforms such as the Stack Overflow (e.g., \cite{shan2025examining}\cite{rao2024generative}\cite{shan2024investigating}), online reviews (e.g., \cite{jia2024leveraging}\cite{shan2025generative}\cite{shan2024generative}). It has also been used to understand users' online search (e.g., \cite{zhao2025shortcuts}) and coding behavior (\cite{shan2024mind}). It offers a novel methodological approach to overcoming traditional research challenges in this area. LLMs like GPT, Gemini, and Claude can generate highly realistic and contextually relevant text. This capability allows for the creation of synthetic virtual survey data under tightly controlled conditions. By systematically manipulating specific variables of interest while holding all others constant, we can conduct comparative experiments that effectively isolate causal effects from the influence of confounding variables often present in observational data \cite{hamalainen_evaluating_2023}.

However, standard LLMs may suffer from factual errors like hallucinations or an inability to process professional information. To overcome these limitations, Retrieval-Augmented Generation (RAG) technology was developed. The RAG workflow typically involves three core steps:
\begin{enumerate}
    \item \textbf{Retrieve:} Upon receiving a prompt, the system does not immediately send it to the LLM. Instead, it uses keywords from the query to search a vast external knowledge base for the most relevant information segments.
    \item \textbf{Augment:} The system then combines these retrieved information segments with the original query to form an augmented prompt, which is richer in content and more contextually specific.
    \item \textbf{Generate:} Finally, this augmented prompt is sent to the LLM. The LLM leverages this additional, reliable context to generate an answer that is more accurate and factually grounded.
\end{enumerate}

Through this mechanism, RAG not only enhances the accuracy and relevance of the generated content but also allows us to "anchor" the model's knowledge to specific, credible data sources. In our research, we can leverage an RAG-LLM framework to create highly realistic and variably controlled experimental materials. For example, based on specific professional skill requirements, we can generate a series of virtual candidate backgrounds that are identical in skills and experience, differing only in the name or other variables to reflect different races or genders.

For the user study, a controlled experimental environment is created using a simulated hiring website, where real recruiters are invited to evaluate a set of constructed freelancer profiles. In these strictly controlled environments, we can precisely measure and isolate the independent impact of specific variables on hiring decisions. This allows for a deeper understanding of the mechanisms behind bias in online markets. This research paradigm, based on user study, holds the promise of overcoming the limitations of traditional methods and opening new avenues for the study of online discrimination.

\section{Methodology}
\subsection{Overall Framework}

In this study, a controlled experiment is employed. Data are obtained through a user study with highly controlled variables.

\begin{figure}
    \centering
    \includegraphics[width=0.99\linewidth]{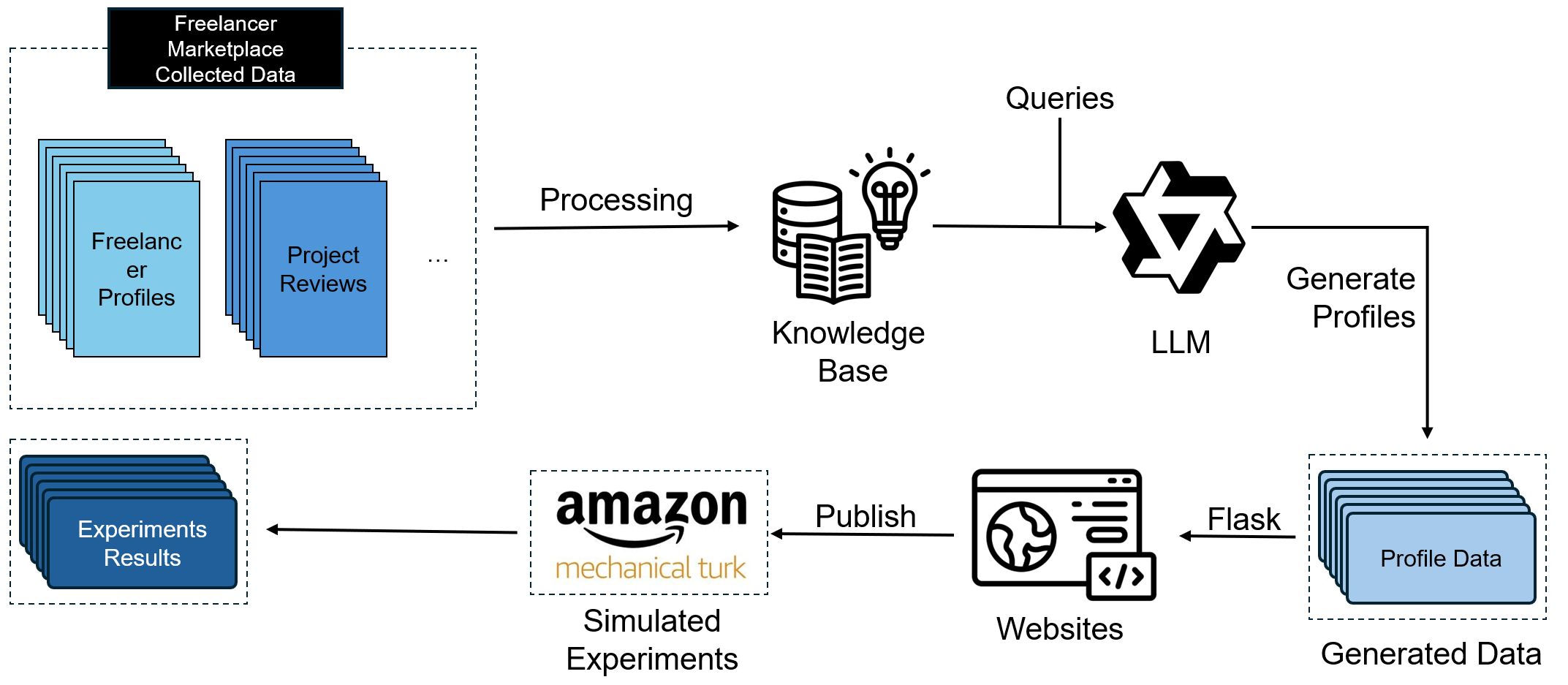}
    \caption{Overall Workflow}
    \label{fig:Overall_Workflow}
\end{figure}

The overall process is shown in Figure \ref{fig:Overall_Workflow}. First, user information is collected from freelancer.com. This information is then processed to build a knowledge base for a Large Language Model (LLM). Subsequently, customized freelancer profiles are generated by the LLM using Retrieval-Augmented Generation (RAG). These profiles are designed to be nearly identical, differing only in the specific variables being investigated. Finally, a webpage is created for the experiment, and participants are recruited through Amazon Mechanical Turk (MTurk) to collect the data.

\subsection{Data Acquisition and Processing}

Scripts are used to collect data from Freelancer.com, a global crowdsourcing marketplace where potential employers post jobs for which freelancers can then bid. This dataset includes project reviews and ratings. The collected profile attributes include details such as Username, User ID, Verification Badges, Overall Rating, and Profile Taglines.

The raw scraped data requires significant preprocessing due to unstructured and inconsistent formatting. A primary challenge is the absence of explicit gender labels on Freelancer.com. To address this, a gender classification model is deployed, which predominantly features self-portraits. Specifically, this pre-trained facial recognition model from Huggingface is employed to analyze these freelancers' avatars, classifying each as "male" or "female" and generating a corresponding gender\_confidence score. To ensure the reliability of the dataset for subsequent analysis, a confidence threshold of 0.75 is applied, and only profiles exceeding this score are included in the final study. For the demographic component, the availability of country data allows for an interaction analysis between geography and gender. This analysis is specifically focused on freelancers from India and the United States, the two most represented nations in the dataset.

\subsection{RAG and LLM}
To generate realistic and contextually relevant freelancer profiles, a Retrieval-Augmented Generation (RAG) pipeline is utilized. The knowledge base for this pipeline is created by vectorizing the processed data with an embedding model from Huggingface. This process embeds the data into a vector space where profiles with similar semantic content are located closer to one another, enabling effective content retrieval.

The core of the content generation model is the Qwen/QwQ-32B Large Language Model (LLM). This model is selected for its advanced text generation capabilities and its proven ability to produce high-quality, consistent output. It is recognized as a powerful open-source solution for this task.

With processed data and a prepared model, virtual information for the experiment is synthetically generated through a multi-stage process to ensure experimental control and anonymity.

The profile generation process begins by processing and vectorizing the collected data to serve as a knowledge base. This knowledge base is then used within a Retrieval-Augmented Generation (RAG) pipeline, which grounds the Large Language Model (LLM) in factual information. In the first step, the most similar profiles from the knowledge base are retrieved. These retrieved documents are then added to the LLM's context as relevant examples. This augmented prompt ensures the final generated profile is not only coherent but also contextually aligned with real-world data. After generation, each profile is manually reviewed and edited to ensure it is high-quality and appropriate for the experiment.

To complete the profiles, names are also synthetically generated using public demographic data. Indian names are sourced from Forebears \cite{noauthor_most_nodate}\cite{noauthor_most_nodate-1}, while American names are created using data from the U.S. Social Security Administration and the Census Bureau \cite{noauthor_top_nodate}\cite{bureau_frequently_nodate}.

\subsection{Website Platform}

With Flask module, interactive webpages for participants on the platform are built and structured into distinct tasks to facilitate precise behavioral data collection. 
\begin{enumerate}
    \item \textbf{Task Selection:} Users begin on a main page, which displays their overall progress and presents two independent experimental tasks: the "Freelancer Comparison Task" and the "Review Comparison Task".

    \item \textbf{Comparison and Voting:} In the Freelancer Comparison Task, the platform displays two freelancer profiles side-by-side and asks the user to vote for the one they would prefer to hire. In the Review Comparison Task, participants are shown relevant review information and are required to answer a series of questions.

    \item \textbf{Task Completion and Survey:} After completing a predetermined number of comparison tasks, users are directed to a completion page. They are then redirected to a final survey to collect demographic information and subjective feedback related to their experience. Attention check questions are added to the survey. If users fail to pass the attention check by choosing the wrong answer of common sense, we exclude the previous selection of this user.
\end{enumerate}

\subsection{Recruit Participants}

Participants are recruited for the study through the MTurk crowd-sourcing platform and directed to our designed experimental website. As tasks are completed, the selections and interaction data from each participant are logged. Following the main tasks, a survey containing several attention check questions is administered. To ensure data integrity, any submissions from participants who fail these checks are filtered out and removed.

\subsection{Statistical Analysis}

The results from the LLM-based user study enable an analysis of the pre-project freelancer selection process. This approach is a notable departure from previous studies, which have been largely confined to analyzing post-project outcomes. With the experiment data, the same analytical methods can be applied to both pre-project and post-project scenarios.

For the statistical analysis, several methods are employed. Star rating data are first binarized into two categories (5-star vs. non-5-star), and a logistic regression model is applied. Logistic regression is also used to analyze count-based outcomes, such as the likelihood of a freelancer being selected. These regression models are specified to include the effects of country and gender, and are analyzed both with and without interaction terms between these variables. Finally, a chi-square test is used to examine the independence between categorical variables.

\section{Results}
\subsection{Methodological Efficiency and Control}
Previous studies are based on traditional observational data and face a critical limitation: the inherent difficulty of controlling for confounding variables. In online marketplace platforms, freelancers' profiles, such as their skills, educational background, and project experience, are typically intertwined with their demographic attributes (e.g., gender, country of origin, or ethnicity). Manually writing each profile is not a feasible alternative. This approach would be time-consuming, with an estimated 10-15 minutes required for each profile. More importantly, it would be extremely difficult to ensure that for every pair of profiles being compared, all attributes are kept highly consistent except for the specific variable under investigation. By contrast, the RAG-LLM framework could address this issue by facilitating precise control and isolation of experimental variables. Rather than analyzing observational data with massive noise, our methodology involves the proactive generation of highly controlled synthetic content for experimentation.

\begin{figure}
    \centering
    \includegraphics[width=0.95\linewidth]{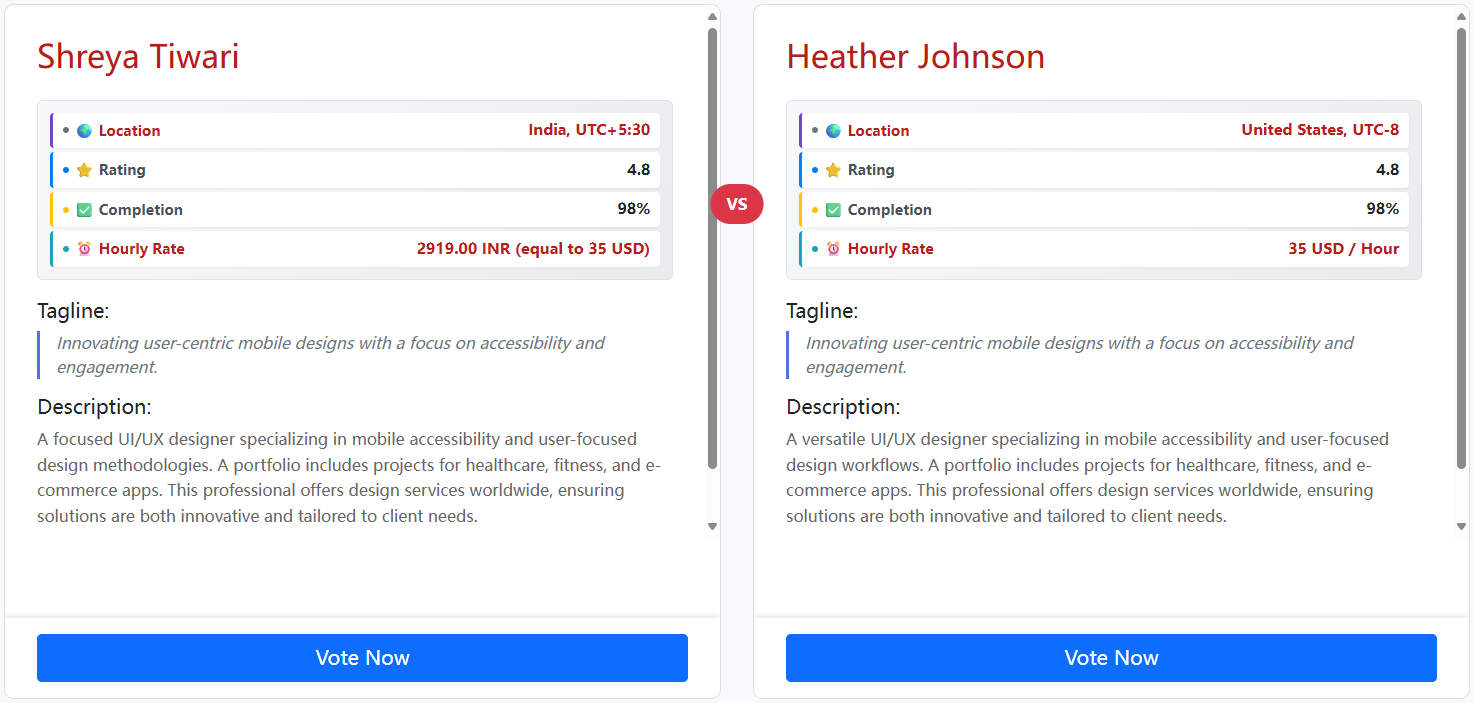}
    \caption{An Example Pair of Freelancers' Profiles}
    \label{fig:Freelancers'_Profiles}
\end{figure}

Figure \ref{fig:Freelancers'_Profiles} shows a screenshot of the web displaying a pair of freelancer profiles for comparison. This particular pair is generated to be nearly identical in qualifications, with the same rating, project completion rate, hourly rate, and tagline. The purpose of this setup is to test for selection bias based on freelancers' demographics, which are signaled through regionally distinct names. The profile descriptions, while not identical, were generated to be closely matched in length and substance. This strategy prevents the experimental setup from appearing unrealistic while ensuring that the content of the descriptions does not become a confounding variable in the participant's decision-making process. 

\subsection{Insights with User Study}
The dataset is collected from the Freelancer website and contains 12,799 freelancer profiles, containing variables such as overall rating, review count, country, and AI-inferred gender.

\begin{figure}
    \centering
    \includegraphics[width=0.9\linewidth]{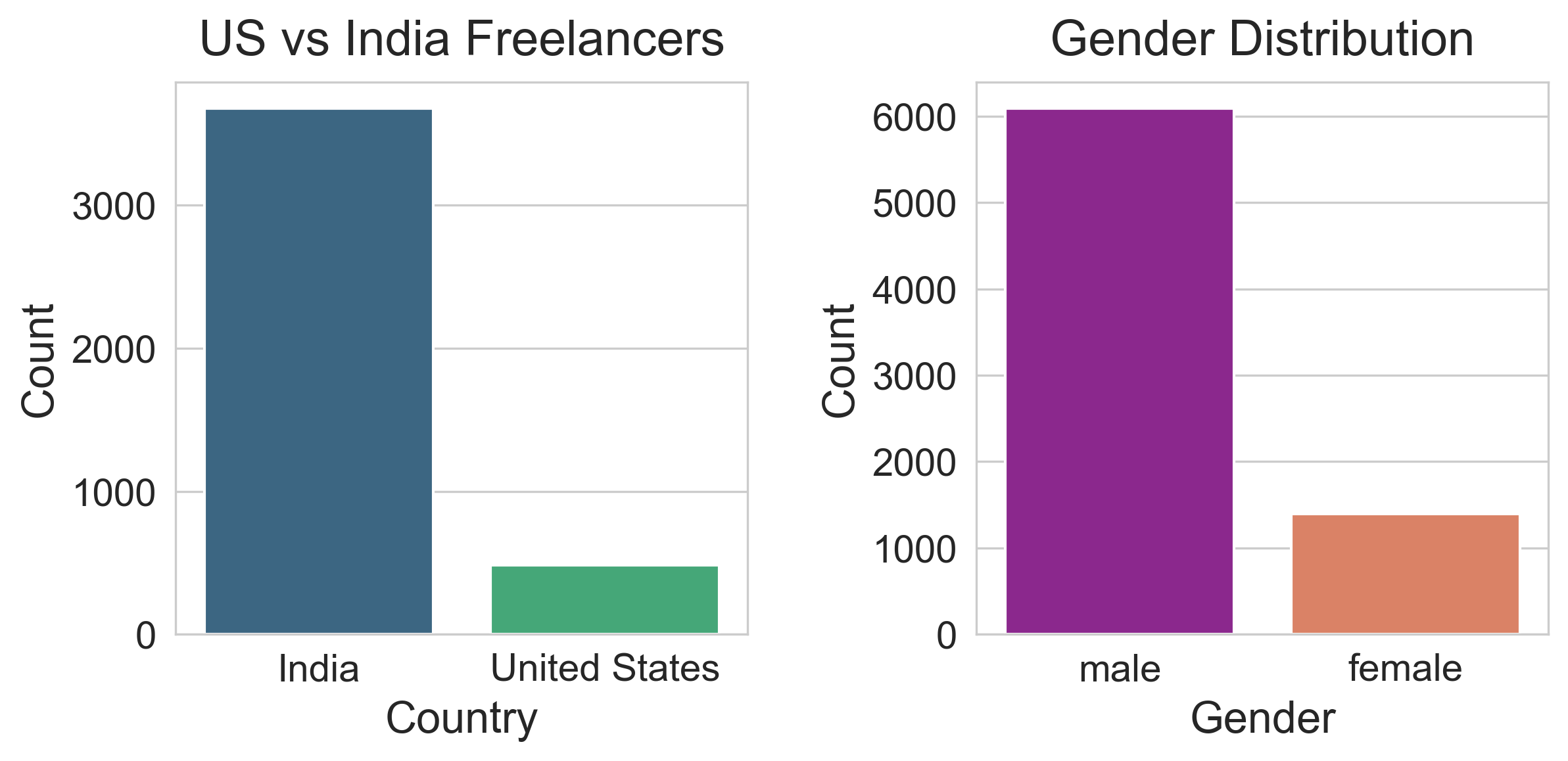}
    \caption{Data Overview from Freelancer.com}
    \label{fig:freelancer_overview}
\end{figure}

Figure \ref{fig:freelancer_overview} provides a visual summary of key distributions within the data. The dataset's composition is further illustrated by comparing the counts of freelancers from our primary countries of interest, the United States and India, and by examining the overall gender distribution. These patterns, particularly the right-skewed distribution of review counts, are consistent with activity data typically found on online marketplaces and social platforms about demographic distribution  \cite{dsouza_freelancing_2024}\cite{noauthor_freelancercom_2024}\cite{noauthor_77_2023}. This analysis sets the stage for the more rigorous statistical modeling used to investigate biases in selection and ratings.

\subsubsection{Hiring Decisions}
A total of 1,980 freelancer profile pairs are generated from the collected dataset to be used in the user study. The data on the display frequency of these pairs and the corresponding selections made by participants are summarized in Table \ref{tab:combined_selection_analysis}.

\begin{table}[H]
\centering
\caption{Pairwise Competition and Win Counts Between Demographic Groups}
\label{tab:combined_selection_analysis}

\begin{subtable}{0.48\textwidth}
    \centering
    \caption{Total Pairwise Competitions}
    \label{tab:selection_appearance_matrix}
    \small
    \begin{tabular}{lcccc}
    \toprule
    \diagbox{\textbf{Group 1}}{\textbf{Group 2}} & \textbf{\makecell{Indian \\ Female}} & \textbf{\makecell{Indian \\ Male}} & \textbf{\makecell{US \\ Female}} & \textbf{\makecell{US \\ Male}} \\
    \midrule
    \textbf{Indian Female} & ---  & 521  & 499  & 0    \\
    \textbf{Indian Male}   & 521  & ---  & 0    & 488  \\
    \textbf{US Female}     & 499  & 0    & ---  & 472  \\
    \textbf{US Male}       & 0    & 488  & 472  & ---  \\
    \midrule
    \textbf{Total}         & 1020 & 1009 & 971  & 960  \\
    \bottomrule
    \end{tabular}
\end{subtable}
\hfill
\begin{subtable}{0.48\textwidth}
    \centering
    \caption{Number of Wins by Group}
    \label{tab:freelancer_win_num_selection}
    \small
    \begin{tabular}{lcccc}
    \toprule
    \diagbox{\textbf{Winner}}{\textbf{Opponent}} & \textbf{\makecell{Indian \\ Female}} & \textbf{\makecell{Indian \\ Male}} & \textbf{\makecell{US \\ Female}} & \textbf{\makecell{US \\ Male}} \\
    \midrule
    \textbf{Indian Female} & --- & 272 & 223 & 0   \\
    \textbf{Indian Male}   & 249 & --- & 0   & 189 \\
    \textbf{US Female}     & 276 & 0   & --- & 245 \\
    \textbf{US Male}       & 0   & 299 & 227 & --- \\
    \bottomrule
    \end{tabular}
    \begin{flushleft}
    \footnotesize
    \textbf{Note:} Each cell represents the number of times the group in the row won a direct competition against the group in the column.
    \end{flushleft}
\end{subtable}
\end{table}

Table~\ref{tab:freelancer_win_prob_selection} and Table~\ref{tab:freelancer_odds_summary} are the results for different demographic groups. There is a strong significant preference for US-based freelancers over their Indian counterparts (p<0.05). While the raw data suggests a slight preference for female freelancers, this tendency is not statistically significant (p>0.3). These results are reflected in the overall winning odds in Table~\ref{tab:freelancer_odds_summary}, where both US groups demonstrate a significant likelihood of winning, while the Indian Male group is significantly more likely to lose.

\begin{table}[H]
\centering
\small
\caption{Freelancer Competition Analysis by Demographic Groups}
\label{tab:freelancer_analysis}

\begin{subtable}{\columnwidth}
\centering
\caption{Winning Probabilities in Competitions}
\label{tab:freelancer_win_prob_selection}
\begin{tabular}{llccc}
\toprule
\textbf{Win} & \textbf{Lose} & \textbf{Odds} & \textbf{95\% CI} & \textbf{P-value} \\
\midrule
\textbf{Indian Male}    & \textbf{US Male}     & 0.633         & [0.527, 0.759] & $<$ 0.001$^{***}$ \\
\textbf{Indian Female}  & \textbf{US Female}   & 0.808         & [0.678, 0.964] & 0.020$^{*}$ \\
\textbf{Indian Female}  & \textbf{Indian Male} & 1.092         & [0.920, 1.297] & 0.335 \\
\textbf{US Female}      & \textbf{US Male}     & 1.079         & [0.901, 1.292] & 0.434 \\
\bottomrule
\end{tabular}
\end{subtable}

\vspace{1em}

\begin{subtable}{\columnwidth}
\centering
\caption{Winning Odds Analysis by Demographic Group}
\label{tab:freelancer_odds_summary}
\begin{tabular}{lccc}
\toprule
\textbf{Group} & \textbf{Odds of Winning} & \textbf{95\% CI} & \textbf{P-value} \\
\midrule
\textbf{Indian Male}    & 0.767         & [0.678, 0.869] & $<$ 0.001$^{***}$ \\
\textbf{Indian Female}  & 0.943         & [0.835, 1.066] & 0.364 \\
\textbf{US Male}        & 1.212         & [1.066, 1.375] & 0.003$^{**}$ \\
\textbf{US Female}      & 1.158         & [1.020, 1.315] & 0.025$^{*}$ \\
\bottomrule
\end{tabular}
\end{subtable}

\begin{flushleft}
\footnotesize
\textbf{Note:} The P-value tests the null hypothesis that the odds ratio is 1 (a 50/50 probability). \\
Significance levels: * $p < 0.05$; ** $p < 0.01$; *** $p < 0.001$
\end{flushleft}
\end{table}

\subsubsection{Leader Decisions}
Parallel to the hiring preference analysis, biases in the perception of leadership qualities among freelancers are shown. This analysis, based on 828 pairwise comparisons where participants selected the more "leader-like" candidate, also reveals a significant demographic bias. The results of these competitions are summarized in Table \ref{tab:combined_leader_analysis}, showing the total number of comparisons between each demographic pair and counts of how many times each group is selected as the leader.

\begin{table}[H]
\centering
\caption{Pairwise Competitions and Selections for Leadership Perception}
\label{tab:combined_leader_analysis}

\begin{subtable}{\columnwidth}
    \centering
    \caption{Total Pairwise Competitions}
    \label{tab:leader_competition_counts}
    \small
    \begin{tabular}{lcccc}
    \toprule
    \diagbox{\textbf{Group 1}}{\textbf{Group 2}} & \textbf{\makecell{Indian \\ Female}} & \textbf{\makecell{Indian \\ Male}} & \textbf{\makecell{US White \\ Female}} & \textbf{\makecell{US White \\ Male}} \\
    \midrule
    \textbf{Indian Female}   & --- & 308 & 315 & 0   \\
    \textbf{Indian Male}     & 308 & --- & 0   & 348 \\
    \textbf{US White Female} & 315 & 0   & --- & 349 \\
    \textbf{US White Male}   & 0   & 348 & 349 & --- \\
    \bottomrule
    \end{tabular}
\end{subtable}
\hfill
\begin{subtable}{\columnwidth}
    \centering
    \caption{Number of Times Selected as Leader}
    \label{tab:leader_win_counts}
    \small
    \begin{tabular}{lcccc}
    \toprule
    \diagbox{\textbf{Winner}}{\textbf{Opponent}} & \textbf{\makecell{Indian \\ Female}} & \textbf{\makecell{Indian \\ Male}} & \textbf{\makecell{US White \\ Female}} & \textbf{\makecell{US White \\ Male}} \\
    \midrule
    \textbf{Indian Female}   & --- & 92  & 76  & 0   \\
    \textbf{Indian Male}     & 68  & --- & 0   & 74  \\
    \textbf{US White Female} & 147 & 0   & --- & 126 \\
    \textbf{US White Male}   & 0   & 149 & 96  & --- \\
    \bottomrule
    \end{tabular}
    \begin{flushleft}
    \footnotesize
    \textbf{Note:} Each cell represents the number of times the group in the row was chosen as the leader.
    \end{flushleft}
\end{subtable}
\end{table}

Table~\ref{tab:combined_leader_odds} shows our findings on demographic bias from two angles. Part (a) details the results from direct, head-to-head comparisons, revealing a strong preference for US candidates over Indian candidates. Part (b) then summarizes the overall performance of each group, confirming that US candidates were chosen far more often (Odds > 1), while Indian candidates were chosen less often (Odds < 1). This shows how the biases in individual matchups lead to the overall group outcomes.

\begin{table}[H]
\centering
\caption{Logistic Regression Analysis of Leadership Selection}
\label{tab:combined_leader_odds}

\begin{subtable}{\columnwidth}
    \centering
    \caption{Winning Odds Ratios in Leadership (Pairwise)}
    \label{tab:leader_win_odds}
    \small
    \begin{tabular}{llccc}
    \toprule
    \textbf{Win Group} & \textbf{Lose Group} & \textbf{Odds} & \textbf{95\% CI} & \textbf{P-value} \\
    \midrule
    \textbf{US Male}       & \textbf{Indian Male}   & 2.014 & [1.524, 2.661] & $<$0.001*** \\
    \textbf{US Female}     & \textbf{Indian Female} & 1.934 & [1.466, 2.551] & $<$0.001*** \\
    \textbf{US Female}     & \textbf{US Male}       & 1.312 & [1.006, 1.712] & 0.051 \\
    \textbf{Indian Female} & \textbf{Indian Male}   & 1.353 & [0.989, 1.851] & 0.069 \\
    \bottomrule
    \end{tabular}
    \begin{flushleft}
    \footnotesize
    \textbf{Note:} The P-value tests the null hypothesis that the odds ratio is 1. \\
    Significance levels: *** $p < 0.001$
    \end{flushleft}
\end{subtable}

\vspace{1em} 

\begin{subtable}{\columnwidth}
    \centering
    \caption{Winning Odds in Leadership Selection (Overall)}
    \label{tab:leader_overall_odds}
    \small
    \begin{tabular}{lccc}
    \toprule
    \textbf{Group}           & \textbf{Odds of Winning} & \textbf{95\% CI} & \textbf{P-value} \\
    \midrule
    \textbf{US White Female} & 1.587 & [1.312, 1.921] & $<$0.001*** \\
    \textbf{US White Male}   & 1.225 & [1.016, 1.477] & 0.037* \\
    \textbf{Indian Female}   & 0.781 & [0.639, 0.956] & 0.019* \\
    \textbf{Indian Male}     & 0.589 & [0.479, 0.725] & $<$0.001*** \\
    \bottomrule
    \end{tabular}
    \begin{flushleft}
    \footnotesize
    \textbf{Note:} The P-value tests the null hypothesis that the odds ratio is 1. \\
    Significance levels: * $p < 0.05$; *** $p < 0.001$
    \end{flushleft}
\end{subtable}
\end{table}

\subsection{Insights with Observational Data}
To complement what we found from the user study, we also adopt traditional methods with scrapping data and analyzing biases in the post-project phase.
\subsubsection{Review Count} 
To analyze the number of reviews received by freelancers, negative binomial regression models are employed to account for the count-based nature of the data. The results of the analysis are presented in Table \ref{tab:combined_review_count}. Without an interaction term, it is shown that freelancers from the United States receive significantly fewer reviews than those from India (IRR = 0.436), while female freelancers receive slightly more reviews than males (IRR = 1.191). For further investigation, an interaction is included, as detailed in Table \ref{tab:combined_review_count}b. A significant interaction effect is found (p = 0.031), which reveals that the influence of gender on review counts depends on the freelancer's country. Specifically, while female freelancers in the Indian group receive about 24\% more reviews than their male counterparts (IRR = 1.237), female freelancers in the U.S. group receive approximately 22\% fewer reviews compared to U.S. males. This finding confirms that the effect of perceived gender on this outcome differs significantly between the two countries.

\begin{table}[H]
\centering
\caption{Results for Review Counts}
\label{tab:combined_review_count}

\begin{subtable}{0.48\textwidth}
    \centering
    \caption{Model without Interaction Term}
    \label{tab:review_count_no_interact_freelancer}
    \small
    \begin{tabular}{lcccc}
    \toprule
    \textbf{Variable} & \textbf{Coeff.} & \textbf{IRR} & \textbf{95\% CI} & \textbf{p-value} \\
    \midrule
    \textbf{Intercept} & 4.737 & 114.07 & [107.9, 120.6] & $<$0.001*** \\
    \textbf{Female}    & 0.175 & 1.191  & [1.055, 1.345] & 0.005** \\
    \textbf{US}        & -0.831& 0.436  & [0.361, 0.526] & $<$0.001*** \\
    \bottomrule
    \end{tabular}
    \begin{flushleft}
    \footnotesize
    \textbf{Notes:} \\
    IRR = Incidence Rate Ratio \\
    Ref. categories: Male; India \\
    ** $p < 0.01$; *** $p < 0.001$
    \end{flushleft}
\end{subtable}
\hfill 
\begin{subtable}{0.48\textwidth}
    \centering
    \caption{Model with Interaction Term}
    \label{tab:review_count_interact_freelancer}
    \small
    \begin{tabular}{lcccc}
    \toprule
    \textbf{Variable} & \textbf{Coeff.} & \textbf{IRR} & \textbf{95\% CI} & \textbf{p-value}\\
    \midrule
    \textbf{Intercept}    & 4.729  & 113.22 & [107.1, 119.7] & $<$0.001***\\
    \textbf{Female}       & 0.213  & 1.237  & [1.089, 1.406] & 0.001***\\
    \textbf{US}           & -0.719 & 0.487  & [0.390, 0.608] & $<$0.001***\\
    \textbf{Female}$\times$\textbf{US} & -0.463 & 0.629  & [0.413, 0.960] & 0.031*\\
    \bottomrule
    \end{tabular}
    \begin{flushleft}
    \footnotesize
    \textbf{Notes:} \\
    Ref. categories: Male; India \\
    * $p < 0.05$; *** $p < 0.001$ \\
    \end{flushleft}
\end{subtable}
\end{table}

\subsubsection{Ratings} 
The rating data are observed to be highly left-skewed because of the high frequency of 5-star reviews. Therefore, for the logistic regression analysis, the ratings are converted into a binary variable: '1' for receiving a non-5-star rating, and '0' for receiving only 5-star ratings.

The results, presented in Table \ref{tab:combined_rating_analysis}, reveal significant demographic effects. The initial model without interactions shows that female freelancers have 51.2\% higher odds of receiving a non-5-star rating compared to their male counterparts (OR = 1.512, p < 0.001). In contrast, freelancers perceived as being from the United States have 37.9\% lower odds of receiving an imperfect rating compared to those from India (OR = 0.621, p = 0.019), indicating they are significantly more likely to receive perfect 5-star scores. The model with interaction includes an interaction term, showing that this interaction is not statistically significant (p = 0.232). This suggests that the tendency for female freelancers to receive lower ratings is consistent across both countries in our dataset.

\begin{table}[H]
\centering
\caption{Results for Receiving a Non-5-Star Rating}
\label{tab:combined_rating_analysis}

\begin{subtable}{0.48\textwidth}
    \centering
    \caption{Model without Interaction Term}
    \label{tab:rating_no_interact_freelancer}
    \small
    \begin{tabular}{lccc}
    \toprule
    \textbf{Variable} & \textbf{Odds Ratio (OR)} & \textbf{95\% CI} & \textbf{p-value} \\
    \midrule
    \textbf{Intercept} & 0.720 & [0.644, 0.805] & $<$0.001*** \\
    \textbf{Female}    & 1.512 & [1.185, 1.930] & $<$0.001*** \\
    \textbf{US}        & 0.621 & [0.418, 0.923] & 0.019* \\
    \bottomrule
    \end{tabular}
    \begin{flushleft}
    \footnotesize
    \textbf{Notes:} \\
    Ref. categories: Male; India \\
    * $p < 0.05$; *** $p < 0.001$
    \end{flushleft}
\end{subtable}
\hfill 
\begin{subtable}{0.48\textwidth}
    \centering
    \caption{Model with Interaction Term}
    \label{tab:rating_interact_freelancer}
    \small
    \begin{tabular}{lccc}
    \toprule
    \textbf{Variable} & \textbf{Odds Ratio (OR)} & \textbf{95\% CI} & \textbf{p-value} \\
    \midrule
    \textbf{Intercept}    & 0.714 & [0.637, 0.799] & $<$0.001*** \\
    \textbf{Female}       & 1.584 & [1.227, 2.045] & $<$0.001*** \\
    \textbf{US}           & 0.726 & [0.457, 1.152] & 0.174 \\
    \textbf{Female}$\times$\textbf{US} & 0.583 & [0.241, 1.412] & 0.232 \\
    \bottomrule
    \end{tabular}
    \begin{flushleft}
    \footnotesize
    \textbf{Notes:} \\
    Ref. categories: Male; India \\
    *** $p < 0.001$
    \end{flushleft}
\end{subtable}
\end{table}

\section{Conclusion}

In this work, a novel method of using a  Large Language Model with Retrieval-Augmented Generation to generate datasets for variable-controlled experiments is employed. In this section, we briefly summarize the key findings obtained through the application of this new approach.

\subsection{Key Findings}
\textbf{From the analysis of user study data, we find:}
\begin{itemize}
    \item In initial hiring selections, U.S. freelancers (of both genders) are significantly more likely to be chosen than their Indian counterparts. Within the same country, however, there is no significant difference in selection rates between genders.
    \item When assessing leadership qualities, a strong preference for U.S. candidates emerges. For example, U.S. males are twice as likely as Indian males to be perceived as leaders. Overall, both U.S. male and female profiles are significant ``winners,'' being chosen as more leader-like, while Indian male and female profiles are significant ``losers.''
\end{itemize}
\textbf{From the analysis of collected data, we find:}
\begin{itemize}
    \item Freelancers from the United States receive significantly fewer reviews than their Indian counterparts, while female freelancers receive slightly more reviews than males.
    \item The effect of gender on review counts is contingent on the country. Specifically, in India, female freelancers receive approximately 24\% more reviews than males, whereas in the U.S., they receive about 22\% fewer reviews than their male counterparts.
    \item Female freelancers are 51.2\% more likely than males to receive non-perfect (i.e., non-5-star) ratings. In contrast, U.S. freelancers are less likely to receive negative feedback and thus more likely to get perfect scores than their Indian peers. The interaction effect between gender and country on ratings was not significant.
\end{itemize}
Previous studies have proven that demographic factors impact freelancer ratings, with a consistent advantage observed for freelancers who are from the U.S. and for those who are male. Our study's findings are consistent with this pattern, providing further confirmation for this argument. Additionally, our analysis of the pre-project stage shows that female freelancers are a little bit more likely to be selected. Simultaneously, the strong advantage for U.S. freelancers persists, even when their profiles are identical to others in terms of skills, pay, and other qualifications.

\subsection{Gender Bias in Different Hiring Stages}

The seemingly contradictory findings on gender bias from the Freelancer.com data and the user study reveal the complexity of how bias operates at different stages.

In our experiment, participants viewed anonymous, AI-generated profiles of similar quality. Their decisions are based solely on first impressions, without any subsequent collaboration or communication. This setup cleanly isolates bias at the initial hiring stage. The finding that gender bias is not significant here suggests that when all other factors (such as skills and portfolio) are standardized, individuals may not make irrational, gender-based choices at the point of hire.

In contrast, the marketplace data allows for the measurement of evaluation bias, defined as the feedback provided to a freelancer after project completion. This reflects not just the first impression of a profile but also the entire project lifecycle, including communication, collaboration, delivery quality, and client satisfaction. The data, which shows that women receive lower ratings (i.e., are more likely to receive non-5-star reviews), indicates that clients may hold stricter, unconscious biases against women during long-term collaboration and final evaluation. They might judge women's work by a different standard or be more prone to forming negative impressions during communication.

This paradox reveals a crucial insight: bias against women may be less pronounced at the hiring stage but more severe during the collaboration and evaluation phases. It paints a picture where women may not struggle to get hired, but are more likely to face unfair judgment in their work.

\subsection{Regional Bias in Different Hiring Stages}

Unlike gender bias, regional bias is highly significant and consistent across both of our analyses, pointing to a more deeply entrenched and pervasive form of prejudice. We propose two potential explanations:

\begin{itemize}
    \item \textbf{In-Group Bias \& Cultural Proximity:} Clients may naturally gravitate towards U.S. freelancers with whom they share a cultural background, language conventions, and compatible time zones.
    \item \textbf{Stereotypes of Developed Countries:} A common cognitive shortcut or stereotype may equate professionals from developed nations like the U.S. with higher quality, professionalism, and reliability. This ``halo effect'' gives U.S. freelancers a significant advantage from the outset, even when their actual skills are comparable to their Indian counterparts.
\end{itemize}

Regional bias appears to be systemic, permeating the entire process from first impressions (selection bias) to post-project reviews (evaluation bias). Clients not only prefer U.S. freelancers at the selection stage but also tend to give them higher ratings after the collaboration is complete.

\section{Usage of Generative AI Statement}
Generative AI (Google Gemini) was used for two specific purposes: (1) to generate LaTeX code for table formatting and (2) to assist with text polishing, including grammar correction and improving sentence clarity. The intellectual content, ideas, and analysis presented are the original work of the author, who takes full responsibility for the final version of the text.


\bibliographystyle{unsrt}
\bibliography{references}


\end{document}